# THE CORE OF CLUSTER-LENSES

*Mass Distribution & Background Galaxies Properties*

J.P. KNEIB

*Institute of Astronomy, Madingley Road, Cambridge CB3 0HA*

AND

G. SOUCAIL

*Observatoire Midi-Pyrénées, 14 Av. E. Belin, 31400 Toulouse*

## 1. Introduction

The confirmation by Soucail et al. (1988) that the nature of the giant arc in A370 is due to gravitational lensing (GL) has brought a new very interesting class of lenses in the GL field: clusters of galaxies. This discovery clearly demonstrated that the core of cluster lenses are very massive and that the mass distribution is peaked, and therefore allows us to see background galaxies distorted but magnified (Fort & Mellier 1994). Modeling of multiple arcs (Mellier et al 1993, Kneib et al 1993) have shown that the mass distribution in the very centre follows the geometrical distribution (ellipticity and orientation) of the stars in the halo of the central cluster galaxy. Total mass estimate derived from X-ray analysis and lensing analysis apparently differs (Miralda & Babul 1995), however recent analysis shows a more complex picture that will lead to better understanding of physical processes in the core of clusters. Knowing the mass distribution of these lenses, we can use them as gravitational telescopes to study the properties of background galaxies and derive a likely distance estimate of faint background galaxies. We use $H_0 = 50h$ km/s/Mpc, $\Omega_0 = 1$ throughout this contribution.

## 2. Known Cluster Lenses

In table 1 we give a non-exhaustive list of clusters in which giant arcs have been discovered (and in a few cases confirmed by spectroscopy). There are 23 such clusters, 11 with known multiple images and 12 with at least one known arc redshift. About 66% of the clusters have a single bright giant elliptical (cD or gE) galaxy at their centre, ∼17% with a dense group of galaxies (g) at the center and ∼17% show bimodal (2gE) or complex morphologies (e.g. cD+gE). For most of them, we give the total mass within the radius of the giant arc. This value comes either from a precise modeling using multiples images (m,★), a simple elliptical modeling (e,□) or a circular modeling (c,↓). Furthermore, if no information exists on the distance of the giant arc, we assume $z_{arc} = 0.8$.

TABLE 1. Cluster-lenses with giant arcs

| Cluster | Type | $z_c$ | $z_{arc}$ | S | $r_{arc}$ (kpc) | $M(< r_{arc})$ ($10^{14}\ M_\odot$) | M |
|---|---|---|---|---|---|---|---|
| PKS0745 | (gE) | 0.103 | 0.433 | * | 45.9 | 0.25 | e |
| MS0955 | (gE) | 0.145 | *0.800* |  | 38.5 | 0.22 | c |
| A2104 | (gE) | 0.155 | *0.800* |  | 31.1 | 0.10 | c |
| A1689 | (g) | 0.170 | *0.800* |  | 161.3 | 2.40 | c |
| A2218a | (cD+gE) | 0.175 | 0.702 | * | 78.5 | 0.50 | m |
| A2218b | (cD+gE) | 0.175 | 1.034 | * | 236.5 | 2.38 | m |
| MS0440 | (g) | 0.190 | 0.530 | * | 90.0 | 0.93 | c |
| A2163 | (gE) | 0.201 | 0.728 | * | 66.1 | 0.75 | c |
| A963 | (gE) | 0.206 | 0.771 | * | 51.7 | 0.35 | c |
| A2219N | (gE) | 0.225 | *1.000* |  | 100.0 | 1.06 | m |
| A1942 | (gE) | 0.226 | *0.800* |  | 37.2 | 0.18 | c |
| A2390 | (gE) | 0.231 | 0.913 | * | 177.0 | 1.60 | m |
| A2397 | (gE) | 0.240 | *0.800* |  | 69.8 | 0.45 | c |
| S295 | (gE) | 0.299 | 0.930 | * | 32.9 | 0.14 | c |
| AC114 | (gE) | 0.310 | 1.860 | * | 56.0 | 0.25 | c |
| MS2137 | (gE) | 0.313 | *1.000* |  | 87.4 | 0.63 | e |
| Cl0500 | (gE) | 0.316 | *0.800* |  | 124.7 | 1.50 | c |
| GHO2154 | (gE) | 0.320 | 0.721 | * | 34.2 | 0.20 | c |
| Cl2244 | (g) | 0.331 | 2.236 | * | 46.5 | 0.20 | c |
| A370 | (2gE) | 0.374 | 0.724 | * | 140.0 | 1.00 | m |
| Cl0024 | (g) | 0.398 | *1.300* |  | 166.7 | 2.00 | e |
| Cl0302 | (gE+E) | 0.424 | *0.800* |  | 122.3 | 1.60 | c |
| Cl2236 | (gE+E) | 0.560 | 1.116 | * | 87.6 | 0.30 | m |
| MS2053 | (gE) | 0.583 | *1.000* |  | 116.8 | 1.50 | c |

In the case of circular model we can write simply: $M(< r_{arc}) = \pi \Sigma_{crit} r_{arc}^2$ with $\Sigma_{crit} = c^2/4G\ D_S D_L/D_{LS}$. We should stress here that this results is independent of the mass profile only if the mass distribution is circular and centered on the brightest cluster galaxy.

However, in most cases the circular model gives an upper mass estimate because of the structure of the mass distribution (elliptical or more complex distribution). Modeling the mass distribution with an elliptical mass distribution generally decreases the mass estimate by 10 to 20%. Figure 1 shows the mass enclosed in the radius of the arc versus its distance to the central cluster galaxy. The lines correspond to isothermal mass distribution with a core of $50h^{-1}$kpc and a velocity dispersion of respectively 1000km/s and 1500km/s. This plot delimits nicely the possible mass profiles, although it does not give a perfect picture because there is some scatter in the total mass from cluster to cluster. Moreover it clearly contradicts the virial masses estimated from cluster velocity dispersion higher than 1500 km/s. This demonstrates that many clusters are still dynamically young and not virialized with ongoing mergers.

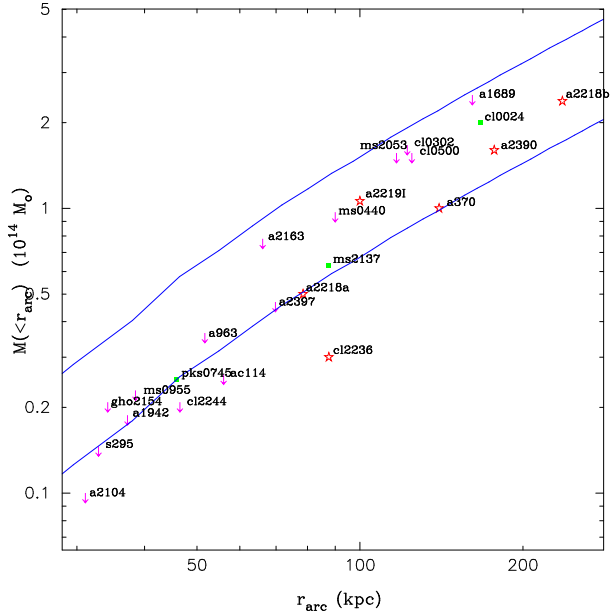

*Figure 1.* Mass within the Einstein radius for clusters of Table 1.

## 3. Mass Distribution

### 3.1. MULTIPLE IMAGES MODELING

The circular model although robust in the mass estimate does not give any clue on the exact mass distribution. The thinness of the arc is the product of the intrinsic size of the arc and the slope of the mass profile at the location of the arc, and therefore shall be analysed carefully. The curvature radius of the arc is sensitive to the "core-radius" and the ellipticity of the mass distribution. In the limit of $r_{curv} \sim r_{arc}$ this gives the constraint $r_{core} < r_{arc}$. Translated to our sample this gives $r_{core} < 100h^{-1}$kpc.

In the case of giant arcs, they are very often multiple-image systems, and a more detailed analysis can be done (e.g. Mellier et al 1993, Kneib et al 1993). Three classes of large arcs are examined:

- fold arc (2 images): either one counter-image (in the case of naked cusp which is likely the case of very elliptical mass distribution) or three counter-images (but one de-amplified) otherwise [e.g. MS2137, Cl2244, A2218, A2219, AC114, Cl0302?, GHO2154?, MS0440?],
- cusp arc (3 images): no-counter images in the naked cusp case or two counter-images (but one de-amplified) otherwise [e.g. A370, Cl0024, S295],
- radial arc (2 images): generally only one counter-image [e.g. MS2137, A370, A2390, AC118].

Finding the multiple images in a cluster is usually an iterative process. High resolution (either with seeing $< 0.7''$ for ground telescope or the use of the HST) or colour matching allows to identify multiple images systems. Once a multiple image system is determined and a mass model can explain it, it is easy to check if other images within the multiple image area have counter-image or not and to use the predictive power of the model.

From the multiple-image system modeling it is now convincing that the total mass distribution follows the orientation and ellipticity of the halo of the brightest cluster galaxies, and that the core radius of the mass distribution is $\sim 50h^{-1}$kpc. The compactness is true in the case of a cluster with a giant elliptical in the center but also in the case where a dense group of galaxies sits in the centre (Cl0024: Smail et al 1995).

More recently, thanks to the high resolution and stability of the HST, a plethora of arcs and arclets have been revealed in the core of some cluster-lenses (A2218, Cl2244, A2390, AC118, AC114, MS0440, MS2137, Cl0024). In the case of A2218, at least 7 sets of multiple images are detected (Kneib et al 1995) which allow a precise modeling of the mass distribution. Furthermore, the HST images suggest that the mass is concentrated around the brightest elliptical galaxies.

3.2. COMPARISON WITH ROSAT/HRI X-RAY MAP

The first comparison of the X-ray map and lens modeling in A370 (Mellier et al 1994) revealed identical morphology, namely two clumps of mass centered on the two brightest ellipticals. However it was difficult to give more precision as the number of counts detected in X-ray was quite low. It is likely however that A370 corresponds to two clusters separated by $\sim 30$Mpc as the relative velocity of the two giant ellipticals suggests it. Miralda-Escudé & Babul (1995) noticed that the total mass estimate in A2218 & A1689 differs by a factor of 2 between lensing and X-ray, although A2163 gives better agreement. Allen et al (1995) made a more precise study of the X-ray gas distribution using both ASCA & ROSAT data and demonstrated with a multi-phase model for the cooling flow cluster PKS0745 that the total mass estimate can agree within a few percents within the arc radius. Deeper ROSAT/HRI images can however tell us more on the substructure of the mass distribution. For example Pierre et al (1995) made a very deep exposure with ROSAT of A2390 and were able to relate a substructure in the X-ray to an enhancement of the total mass as revealed by the presence of the straight arc and an almost "straight" shear field. Further studies will put better constraints on the total mass profile in the central part and probably improve the understanding of physical processes in the core of clusters.

## 4. Background Galaxies Properties

### 4.1. MORPHOLOGY, SIZE, CONTENTS, DYNAMICS OF HIGHLY MAGNIFIED GALAXIES

Smail et al in this proceedings show that the intrinsic sizes of the giant arcs with known redshift are smaller than their present counterpart.
The morphology viewed from HST of the highly magnified arcs display generally complex structure with many knots, but there is at least one case (A2218 #359) where the arc is an elliptical red galaxy.
Pelló et al (1992) first measured a velocity gradient in the straight arc of A2390. Other velocity gradients were detected in Cl2236 (Melnick et al 1993) and PKS0745 (Allen et al 1995). Velocity gradient do probably exists in other arcs (as suggested by their morphology) and are better searched in the most amplified images. Argus/2D spectroscopy mode (specially when used with VLT telescopes) coupled with HST images of giant arcs will allow to have a better understanding of the dynamics of some very distant galaxies.

### 4.2. REDSHIFT DISTRIBUTION OF FAINT GALAXIES

Using the mass model derived from the multiple images with spectroscopically known redshift (therefore absolutely calibrated), it is possible to invert the lens equation for each object in the central part and assign it a most likely redshift. Kneib et al (1994) have first used this method to the arclets in A370, and derived a median redshift $z_{med} \sim 0.8$ for galaxies with intrinsic magnitudes $24 < B_j < 27$. However the largest limitation of this study was probably due to error measurements of the ellipticity of the background galaxies. The refurbished HST will bring a large improvement for this method. Kneib et al (1995) have already applied this inversion to the the shallow exposure of A2218 (3 orbits) and were able to not only to reconstruct the mass of the core of the cluster but constrain the redshift distribution of the background galaxies. Their main results is that the no-evolution prediction (down to R=26) for the background galaxies redshift distribution fits well the observed distortion in A2218.

## 5. Conclusion

The future prospects concerning the core of cluster lenses are:
• the determination of the precise mass distribution down to the galaxy size (i.e. 5–400$h^{-1}$ kpc) especially with the HST images.
• comparison analysis between lensing and X-ray for the "low" redshift cluster (0.15–0.3) for which ROSAT and ASCA data are easily available.
• a better understanding of the physical formation processes of cluster core.
• a better view of high redshift galaxies (morphology, size, content).
• a better constraint on the redshift distribution of faint galaxies by stacking the results from different clusters & check with direct spectroscopy of the brightest arclets.

### Acknowledgment

We thank all our collaborators in this long-term lensing programme, namely B. Fort, Y. Mellier, R. Pelló, R. Ellis, A. Fabian, A. Edge, S. Allen, I. Smail, J. Miralda-Escudé, M. Pierre, H. Börhinger, W. Couch, R. Sharples as well as T. Brainerd, P. Schneider, R. Blandford, M. Rees, for fruitful discussions.

### References

Allen, S.W., Fabian, A.F., Kneib, J.-P. 1995, MNRAS, in press.
Fort, B., Mellier, Y., 1994, Astron. Astr. Rev., 5, 239.
Kneib, J.-P., Mellier, Y., Fort, B., Mathez, G., 1993, A&A, 273, 367.
Kneib, J.-P., Mellier, Y., Fort, B., Soucail, G., Longaretti, P.Y., 1994, A&A, 286, 701.
Kneib, J.-P., Mellier, Y., Pelló, R., Miralda-Escudé, J., Le Borgne, J.-F., Böhringer, H., Picat, J.-P., 1995, A&A, in press.
Kneib, J.-P., Ellis R.S., Smail, I.R., Couch W., Sharples, R., 1995, ApJ, preprint.
Melnick, J., Altieri, B., Gopal-Krishna, Giraud, E., 1993, A&A, 271, L8.
Mellier, Y., Fort, B., Kneib, J.-P., 1993, ApJ, 407, 33.
Mellier, Y., Fort, B., Bonnet, H., Kneib, J.-P., 1994, in "Cosmological Aspects of X-ray clusters of galaxies" NATO ASI series C 441. Waltraut Seitter eds.
Miralda-Escudé, J., Babul, A., 1995, ApJ, in press.
Pierre, M., Le Borgne, J.-F., Soucail, G., Kneib, J.-P., 1995, A&A, preprint.
Smail, I., Dressler, A., Kneib, J.-P., Ellis, R.S., Couch, W.J., Sharples, R.M., Oemler, A., 1995, ApJ, in press.